\documentclass[aps,prb,preprint,groupedaddress,showpacs,amsmath,amssymb]{revtex4}

\usepackage{graphicx}
\usepackage{dcolumn}
\usepackage{bm}

\begin{document}

\title{Hydrostatic and uniaxial pressure dependence of superconducting transition temperature of KFe$_2$As$_2$ single crystals}

\author{Sergey L. Bud'ko$^{1,2}$, Yong Liu$^{1}$, Thomas A. Lograsso$^{1}$, and Paul C. Canfield$^{1,2}$}
\affiliation{$^{1}$Ames Laboratory, US DOE, Iowa State University, Ames, IA 50011, USA}
\affiliation{$^{2}$Department of Physics and Astronomy, Iowa State University, Ames, IA 50011, USA}

\date{\today}

\begin{abstract}
We present heat capacity, $c$-axis thermal expansion and pressure dependent, low field, temperature dependent magnetization for pressures up to $\sim 12$ kbar, data for KFe$_2$As$_2$ single crystals. $T_c$ decreases under pressure with $dT_c/dP \approx -0.10$ K/kbar. The inferred uniaxial, $c$-axis, pressure derivative is positive, $dT_c/dp_c \approx 0.11$ K/kbar. The data are analyzed in comparison with those for overdoped Fe-based superconductors. Arguments are presented that superconductivity in KFe$_2$As$_2$ may be different from the other overdoped, Fe-based materials in the 122 family.

\end{abstract}


\maketitle

\section{Introduction}

Since the recent discovery of superconductivity at elevated temperature in F-substituted LaFeAsO, \cite{kam08a} a large amount of experimental and theoretical effort was concentrated on studies of magnetic, superconducting and normal state properties of Fe - based superconductors and related materials. \cite{joh10a,ste11a,joh11a} Of several families of Fe-based superconductors discovered by now, the 122, AEFe$_2$As$_2$ (AE = alkaline earth and Eu) family, appears to be the most studied one \cite{can10a,nin11a,man10a}  due to the reasonable ease of growing large, high quality, single crystals, the availability of multiple substitution sites and the simplicity of the crystal structure. In the 122 system, superconductivity, with a $T_c \approx 38$ K,  was first reported on K-doping of BaFe$_2$As$_2$. \cite{rot08a}  Subsequently the complete (Ba$_{1-x}$K$_x$)Fe$_2$As$_2$ solid solution series was studied \cite{rot08b, che09a,avc12a} in detail. 

KFe$_2$As$_2$ stands out among the members of the (Ba$_{1-x}$K$_x$)Fe$_2$As$_2$ series, and the Fe-based superconductors in general, as a unique material. It is a stoichiometric end-member of the (Ba$_{1-x}$K$_x$)Fe$_2$As$_2$ series, and a rare example of a stoichiometric Fe-based superconductor, albeit with a rather low $T_c \approx 3.8$ K. \cite{rot08b,che09a,avc12a,sas08a} The in-plane resistivity has a metallic behavior with a remarkable residual resistivity ratio, exceeding 1000 in the best single crystals. \cite{has10a} The reported Fermi surface of KFe$_2$As$_2$ differs from that of the optimally doped  (Ba$_{1-x}$K$_x$)Fe$_2$As$_2$ having three hole pockets, two centered at the $\Gamma$ point in the Brillouin zone, and one around the $M$ point \cite{sat09a} with no electron pockets. Existence of quantum criticality, and nodal or $d$-wave superconductivity was suggested and discussed in number of publications. \cite{has10a,don10a,ter10a,don10b,rei12a,mai12a,tho11a} 

Given the unusual properties of this material, it is of importance to have a broad set of data on its physical properties, in particular, related to its superconducting state. In this work we study the response of KFe$_2$As$_2$'s superconducting transition temperature to external pressure, both hydrostatic, via direct measurements, and uniaxial, that is inferred by using the Ehrenfest relation for the second order phase transition. Hydrostatic pressure effects historically have been studied for many superconductors, including some Fe-based materials. \cite{chu09a,can09a,sef11a} Uniaxial pressure effects are rarely measured directly, \cite{wel92a,tor09a} due to significant technical difficulties, however evaluations of the uniaxial pressure derivatives, $dT_c/dp_i$, ($ i = a, b, c$) by combining the thermal expansion and heat capacity data, have been recently made for several BaFe$_2$As$_2$ - based materials \cite{bud09a,har09a,luz09a,mei12a,boh12a} yielding in-plane and $c$-axis uniaxial pressure derivatives of the opposite signs and distinct evolution of these derivatives with doping.

\section{Experimental details}

KFe$_2$As$_2$ single crystals were grown  by using KAs flux method. \cite{kih10a} Briefly, K (ingot), As (lump), and Fe (powder) were weighed at an atomic ratio of K : Fe : As = 5 : 2 : 6, and loaded into an alumina crucible. A sealing technique with liquid tin melt was developed to suppress the evaporation of K and As elements. \cite{liu12a} All the steps were performed in a glove box under argon gas atmosphere. The sealed fused silica ampule was heated up to 920$^\circ$ C, and then slowly cooled down to 620$^\circ$ C at a rate of 1$^\circ$ C/h. Thin, plate-like crystals, with dimensions up to $8 \times 5 \times 0.2$ mm$^3$,  can be easily mechanically separated from the KAs flux.. The further details of the crystal growth will be reported elsewhere. \cite{liu12a} The crystals are plate-like with the $c$-axis perpendicular to the plate. They are soft, micaceous and moderately air - sensitive. Low-field dc magnetization under pressure, was measured in a Quantum Design Magnetic Property Measurement System, MPMS-5, SQUID magnetometer using a  a commercial, HMD, Be-Cu piston-cylinder pressure cell. \cite{hmd}  Daphne oil 7373 was used as a pressure medium and superconducting Pb as a low-temperature pressure gauge. \cite{eil81a}. Thermal expansion data were obtained using a capacitive dilatometer constructed of OFHC copper; a detailed description of the dilatometer is presented elsewhere. \cite{sch06a} The dilatometer was mounted in a Quantum Design Physical Property Measurement System, PPMS-14, instrument and was operated over a temperature range of 1.8 - 305 K. Due to the morphology of the crystals, the dilation was measured only along the $c$-axis. The heat capacity was measured using a hybrid adiabatic relaxation technique of the heat capacity option in a Quantum Design, PPMS-9 instrument.

\section{Results}

\begin{figure}
\begin{center}
\includegraphics[angle=0,width=120mm]{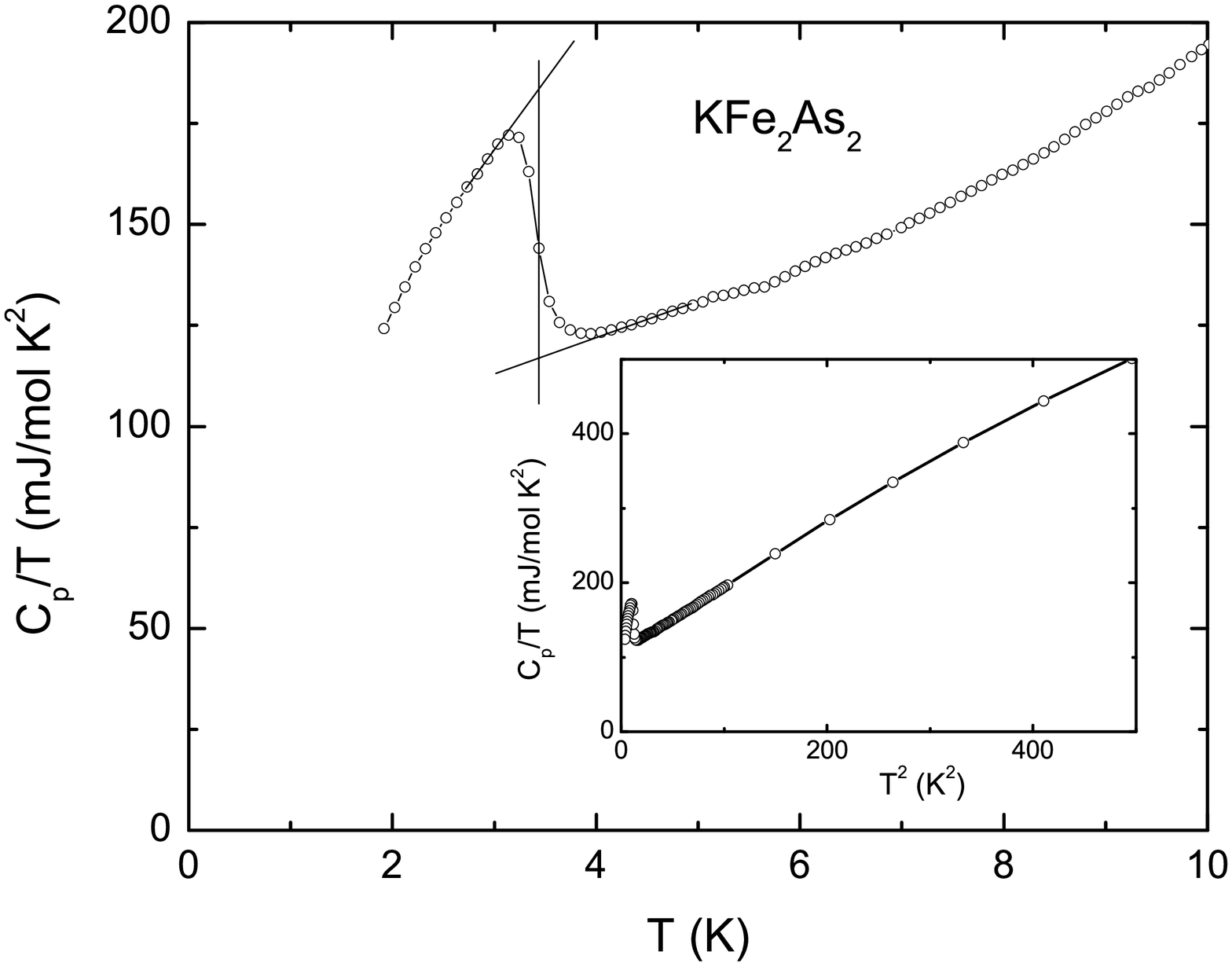}
\end{center}
\caption{ Temperature-dependent heat capacity plotted as $C_p/T$ vs $T$. Lines show how the $\Delta C_p/T|_{T_c}$ was determined. Inset: $C_p/T$ vs $T^2$.  } \label{F2}
\end{figure}

The low temperature heat capacity of KFe$_2$As$_2$ is plotted as $C_p/T$ vs (T) in Fig. \ref{F2}. The jump in specific heat associated with the superconducting transition is rather sharp. From the isoentropic construct, shown in Fig. \ref{F2}, $T_c \approx 3.4$ K,  $\Delta C_p/T|_{T_c} \sim 67$ mJ/mol K$^2$. From the linear fit of the  $C_p/T$ vs $T^2$ above the superconducting transition (Fig. \ref{F2}, inset), the Sommerfeld coefficient, $\gamma \approx 107$ mJ/mol K$^2$, and the Debye temperature, $\Theta_D \approx 224$ K can be estimated. These results are comparable to the literature data. \cite{fuk11a,kim11a,abd12a}

\begin{figure}
\begin{center}
\includegraphics[angle=0,width=120mm]{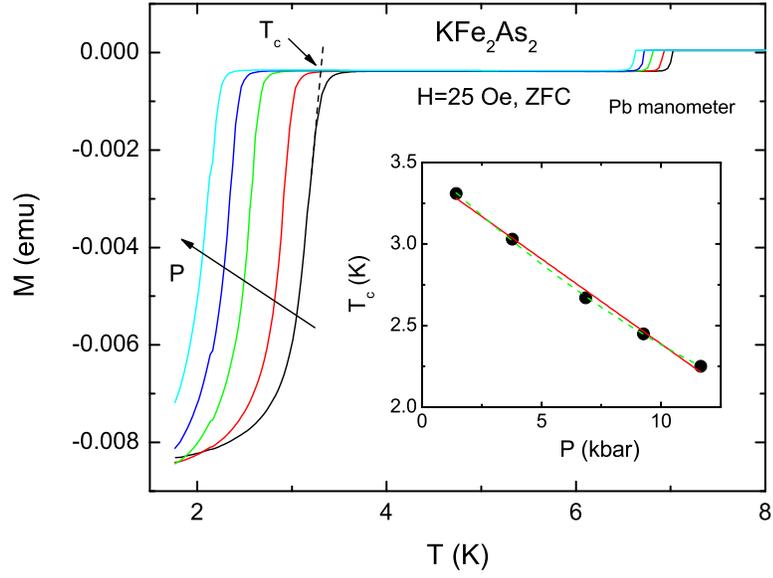}
\end{center}
\caption{(Color online) ZFC temperature-dependent magnetization of KFe$_2$As$_2$ under pressure. The $T_c$ criterion used in the paper is shown for the lowest pressure curve. The signal from the Pb manometer is also shown. Inset - pressure dependence of $T_c$ of KFe$_2$As$_2$. Lines: linear (solid) and second order polynomial (dashed) fit of $T_c(P)$. } \label{F1}
\end{figure}

Low field ($H = 25$ Oe) zero-field-cooled (ZFC) temperature-dependent magnetization data taken under pressures up to $\approx 11.7$ kbar are shown in Fig. \ref{F1}. The signal associated with the superconducting transition in KFe$_2$As$_2$ is sharp, under pressure the transition shifts to lower temperatures without any significant broadening. The pressure dependence of $T_c$ (defined as an onset of transition in magnetization, see Fig. \ref{F2}) is shown in the inset to Fig. \ref{F1}. A linear fit of $T_c(P)$ results in the value of the pressure derivative, $dT_c/dP = - 0.11 \pm 0.01$ K/kbar, so that in a simple, linear, approximation superconductivity in KFe$_2$As$_2$ can be suppressed by a pressure of $\sim 30$ kbar. It appears that the measured $T_c(P)$ dependence has a slight upward curvature. Indeed, one can also fit $T(P)$ with a second order polynomial (shown as a dashed line in the inset to Fig. \ref{F1}). Then the linear coefficient corresponding to the initial $dT_c/dP$ is $- 0.14 \pm 0.01$ K/kbar (the quadratic coefficient equals to $0.0027 \pm 0.0004$ K/kbar$^2$). Very similar results are obtained if a different criterion (e.g. maximum of $dM/dT$) is used.

\begin{figure}
\begin{center}
\includegraphics[angle=0,width=120mm]{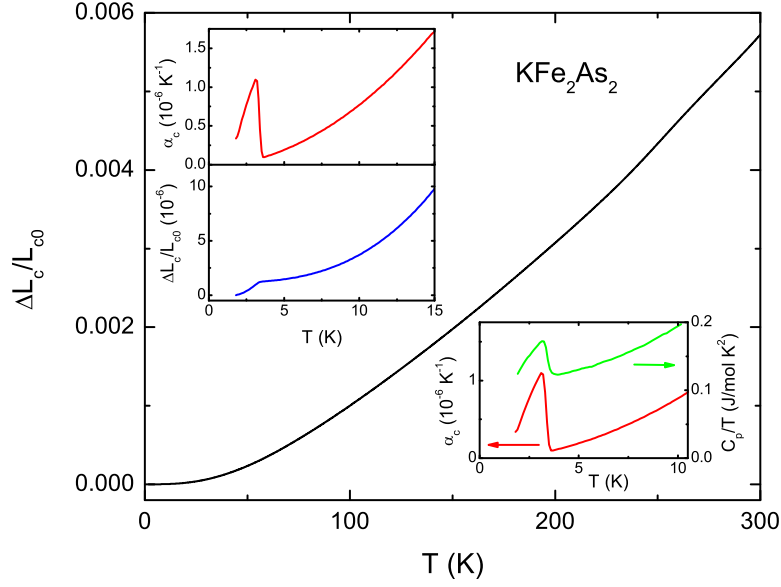}
\end{center}
\caption{(Color online) Temperature-dependent $c$-axis dilation of KFe$_2$As$_2$. The data are normalized to $L_{c0}$ value at 1.8 K. Upper left inset: low temperature $c$-axis dilation and thermal expansion coefficient, $\alpha_c$, with the anomalies at the superconducting transition.  Lower right inset: low temperature thermal expansion coefficient, $\alpha_c$, and heat capacity, $C_p/T$, both showing jump at the superconducting transition. } \label{F3}
\end{figure}

Temperature dependent $c$ - axis dilation, normalized to the value at 1.8 K ($|Delta L_c/L_{c0}$), is shown in Fig. \ref{F3}. The overall behavior is monotonic with some flattening at low temperatures. The relative change in the $c$ - axis from 1.8 to 300 K is similar to that measured in pure AEFe$_2$As$_2$ (AE = Ba, Sr, Ca) \cite{bud10a} and Co - doped BaFe$_2$As$_4$ \cite{bud09a,luz09a}, with the value of the thermal expansion coefficient, $\alpha_c = d(\Delta L_c/L_{c0})/dT$, at room temperature being under $3 \cdot 10^{-5}$ K$^{-1}$. Both, the $c$-axis dilation and the thermal expansion coefficient show a clear anomaly at the superconducting transition.   The jump in the thermal expansion coefficient at $T_c$  (using the criterion graphically similar to that for $\Delta C_p/T|_{T_c}$ in Fig. \ref{F2}) can be estimated as $\Delta \alpha_c|_{T_c} \approx 1.2 \cdot 10^{-6}$ K$^{-1}$. The shape and sharpness of the features in the thermal expansion coefficient and the heat capacity at $T_c$ are very similar (Fig. \ref{F3}, inset), that rationalizes use of the same criterion for both measurements.

\section{Discussion and Summary}

The value of the hydrostatic pressure derivative, $dT_c/dP = - 0.10$ K/kbar, measured for KFe$_2$As$_2$ is rather large. It is close to, but somewhat smaller than $dT_c/dP = - 0.132$ K/kbar reported for stoichiometric, polycrystalline, RbFe$_2$As$_2$. \cite{she12a}  It has to be noted that for slightly overdoped (Ba$_{0.55}$K$_{0.45}$)Fe$_2$As$_2$ rather large, negative, pressure derivatives, $dT_c^{onset}/dP = - 0.15$ K/kbar and  $dT_c^{offset}/dP = - 0.21$ K/kbar were reported as well. \cite{tor08a} The evolution of the $dT_c/dP$ values with K-doping in the (Ba$_{1-x}$K$_x$)Fe$_2$As$_2$ series, in particular in the overdoped region, will be useful for further understanding of superconductivity in these materials. 

The uniaxial, $c$ - axis, pressure derivative of $T_c$ can be inferred using the Ehrenfest relation \cite{bar99a} for the second order phase transition:  $dT_c/dp_i = \frac{V_m \Delta \alpha_i T_c}{\Delta C_p}$, where $V_m$ is the molar volume ($V_m \approx 6.08 \cdot 10^{-5}$ m$^3$ for KaFe$_2$As$_2$, using the lattice parameters \cite{avc12a} at 1.7 K), $\Delta \alpha_i~ (i = a, c)$ is the jump in the thermal expansion coefficient at the phase transition, and $\Delta C_p$ is the corresponding jump in the heat capacity. Using the experimental data above, for KFe$_2$As$_2$ we found out that the $c$ - axis uniaxial pressure derivative is positive, $dT_c/dp_c \approx 0.11$ K/kbar. Since the hydrostatic pressure derivative in this case can be written as $dT_c/dP = 2 dT_c/dp_a + dT_c/dp_c$, we can infer that the uniaxial, $a$-axis (or in-plane) pressure derivative is negative, $dT_c/dp_a \approx - 0.11$ K/kbar. So it appears that KFe$_2$As$_2$ is equally sensitive to uniaxial pressure applied in the $ab$ - plane and along the $c$ - axis, however $T_c$ increases when the pressure is applied along the $c$-axis and decreases when it is applied in the $ab$ - plane. Uniaxial pressure derivatives of $T_c$ in  members of the 122 family were also  reported to have different signs for $ab$-plane and $c$-axis pressure. We are not aware of such data in the (Ba$_{1-x}$K$_x$)Fe$_2$As$_2$ series, but for  Co-doped (electron doping) and P-doped (isoelectronic substitution) BaFe$_2$As$_2$ samples in the overdoped region of the superconducting dome  negative values of $dT_c/dp_c$  and positive values of  $dT_c/dp_{ab}$ were reported. \cite{bud09a,har09a,luz09a,boh12a} If we consider KFe$_2$As$_2$ as an extreme of the overdoped  (Ba$_{1-x}$K$_x$)Fe$_2$As$_2$, it is clearly different, in this respect, from the other (studied so far) overdoped Fe - based superconductors from the 122 family.

It has to be mentioned that the anisotropy and  different signs of the uniaxial pressure derivatives of $T_c$, as inferred from thermal expansion and heat capacity data using the Ehrenfest relation, can shed some light on the differences observed in the values of the critical pressure and position of the superconducting dome that vary in publications describing  the pressure - temperature phase diagrams in the AEFe$_2$As$_2$ materials. \cite{col09a,kot09a,bra10a,dun10a} In many cases of the different anvils - based pressure cells an additional, axial, component of pressure is present. With the plate -  like  AEFe$_2$As$_2$ samples in a "convenient geometry", this means that the $c$-axis pressure in such cells is  slightly higher than the pressure in the $ab$ - plane. Then, for the samples with $dT_c/dp_c < 0$ the measured $dT_c/dP$ is smaller than the truly hydrostatic value, and vise versa, if $dT_c/dp_c > 0$, larger than hydrostatic value of $dT_c/dP$ is observed in an experiment.

\begin{figure}
\begin{center}
\includegraphics[angle=0,width=120mm]{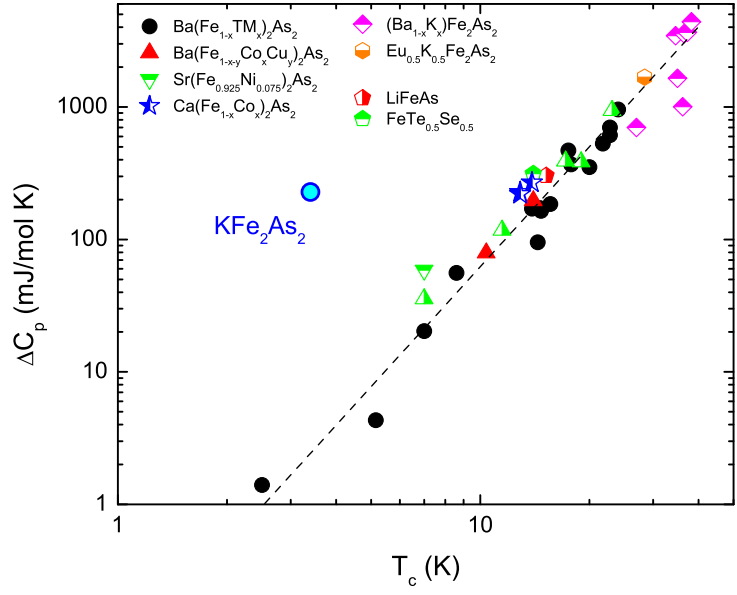}
\end{center}
\caption{(Color online) $\Delta C_p$ vs $T_c$  for the KFe$_2$As$_2$ sample, plotted together with literature data for various FeAs-based superconducting materials. Updated plot \cite{ran12a} is used to show the literature data. The line corresponds to $\Delta C_p \propto T_c^3$.} \label{F4}
\end{figure}

Another clear difference between this material and other Fe - based, 122 superconductors is that the jump in heat capacity at $T_c$ clearly deviates from the trend suggested in Ref. \onlinecite{bud09b} and expanded in Ref. \onlinecite{kim11b}, the so-called BNC scaling (Fig. \ref{F4}). At the same time, our data yield $\frac{\Delta C_p}{\gamma T_c} \approx 0.62$ that is significantly smaller than the 1.43 value expected for BCS superconductors. These observations are consistent with previous publications \cite{kim11a,abd12a} and suggest that superconductivity in  KFe$_2$As$_2$  neither is a conventional BCS, nor should it be considered as just significantly overdoped Fe-based superconductor from the 122 family.
\\

To summarize, the superconducting transition temperature of KFe$_2$As$_2$ decreases under hydrostatic pressure rather fast, with $dT_c/dP \approx - 0.104$ K/kbar. The uniaxial, $c$ - axis, pressure derivative inferred from the thermal expansion and heat capacity measurements, is positive, that differs from the existing data for overdoped Fe-based 122 superconductors. The jump in heat capacity at $T_c$ for KFe$_2$As$_2$ deviates significantly from the empirical trend, $\Delta C_p \propto T_c^3$ (and its recent modifications), observed in a number of Fe-based superconductors. Taken together, these results suggest that the superconductivity in KFe$_2$As$_2$ could be different from just an extreme overdoped case in the 122 family. Detailed, comprehensive studies of the  (Ba$_{1-x}$K$_x$)Fe$_2$As$_2$ series with $x$ approaching 1, once homogeneous single crystalline samples for these intermediate concentrations are available, might shed a light on physics of this interesting material.

\begin{acknowledgments}

Work at the Ames Laboratory was supported by the US Department of Energy, Basic Energy Sciences, Division of Materials Sciences and Engineering under Contract No. DE-AC02-07CH11358. S.L.B. acknowledges partial support from the State of Iowa through Iowa State University.

\end{acknowledgments}

\clearpage

\clearpage

\clearpage

\clearpage

\end{document}